\documentclass[12pt]{article}
\usepackage{graphicx}
\usepackage{amssymb}
\usepackage{epsfig}
\usepackage{amsmath}
\newcommand{\be}{\begin{equation}}
\newcommand{\ee}{\end{equation}}

\newcommand{\LA}{\langle}
\newcommand{\RA}{\rangle}

\newcommand{\bea}{\begin{eqnarray}}
\newcommand{\eea}{\end{eqnarray}}

\newcommand{\La}{\lambda^{a}}
\newcommand{\Lla}{\tilde{\lambda}^{\dot{a}}}
\newcommand{\PP}{\sum_i\La_i\Lla_i}

\newcommand{\dd}{\delta^4}
\newcommand{\la}{\lambda}
\newcommand{\lla}{\tilde{\lambda}}

\DeclareSymbolFont{AMSa}{U}{msa}{m}{n}
\DeclareSymbolFont{AMSb}{U}{msb}{m}{n}
\DeclareMathSymbol{\fieldR}{\mathalpha}{AMSb}{"52}

\newif\ifpdf
\ifx\pdfoutput\uundefined
\pdffalse 
\else
\pdfoutput=1 
\pdftrue
\fi

\begin{document}
\begin{titlepage}
\begin{center}
\hfill {\tt YITP-SB-04-24}\\
\hfill {\tt hep-th/0405086}\\
\vskip 20mm

{\Large {\bf A Note on Twistor Gravity Amplitudes}}

\vskip 10mm

{\bf Simone
Giombi$^{\dagger}$\footnote{sgiombi@grad.physics.sunysb.edu},
Riccardo Ricci$^{\flat}$\footnote{rricci@insti.physics.sunysb.edu},\\
Daniel
Robles-Llana$^{\flat}$\footnote{daniel@insti.physics.sunysb.edu}
and Diego Trancanelli$^{\dagger}$\footnote{trancane@grad.physics.sunysb.edu} }

\vskip 4mm {\em $^{\flat}$ C. N. Yang Institute for Theoretical
Physics,}\\
{\em $^{\dagger}$ Department of Physics and Astronomy}\\
{\em State University of New York at Stony Brook}\\
{\em Stony Brook, NY 11794-3840, USA}\\
[2mm]

\end{center}

\vskip 1in

\begin{center} {\bf ABSTRACT }\end{center}
\begin{quotation}
\noindent

In a recent paper, Witten proposed a surprising
connection between perturbative gauge theory and a
certain topological model in twistor space. In particular,
 he showed that gluon amplitudes are localized on holomorphic 
curves. In this note we present some preliminary considerations on the
possibility of having a similar localization for gravity amplitudes.

\end{quotation}

\vfill
\flushleft{May 2004}

\end{titlepage}

\eject

\section{Introduction}
\setcounter{equation}{0}

Recently Witten found a remarkable connection between
perturbative $\mathcal{N}=4$ Super Yang-Mills theory and the
topological $B$ model on the super
Calabi-Yau space $CP^{3|4}$ \cite{witten} \footnote{Recent
related works can be found in
\cite{MHV, parity, roiban, berkovits, vafa, gukov, ci, kohze, siegel}.}.
Interpreting perturbative amplitudes
in terms of a $D$-instanton expansion in the topological theory, the
conjecture offers a deeper understanding of well-known field
theory results. At tree level, after stripping out the color
information, Yang-Mills theory is effectively supersymmetric and
therefore Witten proposal provides a new, suggestive
approach to study YM amplitudes. In particular some seemingly accidental
properties of scattering amplitudes, like the holomorphicity \footnote{Up to the delta-function of momentum conservation.} of
the MHV Parke-Taylor
formula \cite{parke}
\begin{equation}
\mathcal{C}(1+,\cdot \cdot \cdot ,p-,\cdot \cdot \cdot ,q-,\cdot
\cdot \cdot
,n+)=ig^{n-2}(2\pi )^{4}\delta ^{4}\left( \sum_{i}\lambda _{i}^{a}\tilde{%
\lambda}_{i}^{\dot{a}}\right) \frac{\langle \lambda _{p},\lambda _{q}\rangle
^{4}}{\prod_{i=1}^{n}\langle \lambda _{i},\lambda _{i+1}\rangle },
\label{05}
\end{equation}
receive a new elegant interpretation in terms of localization over
certain subloci of the target space $CP^{3|4}.$ More precisely,
according to the conjecture the $l$ loop contribution to the
$\mathcal{N}=4$ SYM $n$ gluon scattering amplitude is localized in
twistor space on an algebraic curve of degree and genus given by
\bea
& &d=q-1+l\nonumber\\
& &g\leq l \label{1} \eea where $q$ is the number of negative
helicity external legs.

For instance, the holomorphicity of (\ref{05})  allows to check
that the  MHV amplitudes, once transformed to twistor space,
 are indeed supported on $d=1$ genus zero
curves in $CP^{3}$ (the body of the
supermanifold $CP^{3|4}$)
\begin{equation}
\tilde{\mathcal{C}}(\lambda _{i},\mu _{i})=ig^{n-2}\int
d^{4}x\prod_{i=1}^{n}\delta ^{2}(\mu
_{i\dot{a}}+x_{a\dot{a}}\lambda _{i}^{a})f({\lambda _{i}}).
\end{equation}

\textit{A priori} one would expect a tree YM amplitude with $q$
negative gluons to receive contributions not only from $d=q-1$
genus zero curves but also from all possible decompositions in
disconnected curves $C_i$ of degree $d_i$ such that $\sum_{i}d_i=q-1$.
An explicit calculation of the connected contribution to all the
googly amplitudes $\mathcal{C}(+,+,-,-,-)$ has been performed in
\cite{roiban} by
integrating over the moduli space of connected curves with genus
zero and degree 2. Surprisingly the result correctly
reproduces the known amplitudes \textit{without} any additional
contribution from disconnected configurations.

In \cite{MHV} the limit of totally disconnected configuration,
that is $q-1$ curves of degree 1, has been considered. A
particular class of Feynman diagrams (MHV tree diagrams) was
built in which each vertex corresponds to a $d=1$ genus zero curve
and the contribution of each vertex is the MHV
$\mathcal{C}_n(-,-,+,\ldots,+)$ amplitude suitably
extended off-shell. The vertices are joined using the scalar
propagator $1/p^2$. Quite amazingly this set of totally
disconnected configurations is also enough to reproduce all the
googly amplitudes and likely all the tree YM amplitudes \cite{MHV}, \cite{ci}.
On the string theory side, the advantage of the disconnected
prescription is that we can
avoid integrating over the moduli space of connected curves and
therefore drastically simplify the task of computing tree YM
amplitudes. On the other hand, from the field theory point of view,
 the simplicity of
the MHV prescription offers a very efficient way to calculate
multi-gluon tree amplitudes \footnote{An explicit example of the power of this method has been given in
\cite{MHV}, where a simpler form of $\mathcal{C}_n(-,-,-,+,\ldots,+)$, previously computed in \cite{kosow}, was obtained.}. A
proof \footnote{ Modulo subtleties regarding the choice of the
contour of integration.} of the equivalence of connected and
disconnected prescriptions has been presented in \cite{gukov}. The
MHV formalism has been also successfully extended to YM coupled to
fundamental fermions \cite{kohze}.

In this note we present some preliminary considerations on gravity
amplitudes following some suggestions in \cite{witten}. The closed
string sector of the $B$ model on $CP^{3|4}$ should presumably
describe $\mathcal{N}=4$ conformal supergravity, which at tree
level reduces to conformal gravity. Ordinary gravity amplitudes
would be related not to the closed sector of the $B$ model on
$CP^{3|4}$ but to that of a yet unknown topological twistor string
theory which probably describes $\mathcal{N}=8$ supergravity. Even
though the correct framework for studying gravity has not been
established, some preliminary indications on localization of tree
level gravity amplitudes can be given. Some initial analysis of
the MHV case was already given in \cite{witten}. The crucial
difference with respect to YM is that the $n$ graviton MHV
amplitude is \textit{not holomorphic} in the spinor helicity
variables in Minkowski space. This non holomorphic dependence is
nonetheless very simple, namely polynomial. The polynomial
dependence implies that MHV gravity amplitudes are supported again
on $d=1$ curves, but now with a multiple derivative of a
delta-function, as we review in the next Section.

It is natural to investigate if this behavior
persists for non-MHV cases. In Section 2 we check
the simplest non trivial case, namely the googly amplitude
$\mathcal{M}(+,+,-,-,-)$. Constructing a suitable
differential operator which annihilates
the amplitude, we verify that this  is supported on a connected degree 2
curve of genus zero. This is similar to what happens for
the corresponding googly YM amplitude, with the difference that we
now have a derivative of a delta-function support.

This does not exclude \textit{a priori} the presence of
disconnected contributions. In Section 3 we comment on the
possibility of a MHV decomposition of gravity amplitudes. Note
that even without knowing the underlying string theory, having a
MHV-like diagrammatic expansion would dramatically simplify the
calculation of gravity amplitudes, which are notoriously
complicated and in many cases not known in closed form.

The vertices are built using the MHV prescription for YM and the
KLT relations,  which in general express closed string amplitudes
as a sum of products of open string amplitudes, in the field
theory limit \cite{klt}. Differently from the gauge theory case it
is not possible to construct MHV gravity diagrams using only
holomorphic vertices.  The only diagrams which can be built using
holomorphic vertices correspond to  amplitudes of the form
$\mathcal{M}_n(+,-,\ldots ,-)$. As in YM these are known
to vanish. Using the completely disconnected prescription we
verify that  the  MHV diagrams for $\mathcal{M}(+,-,-,-)$ and
$\mathcal{M}(+,-,-,-,-)$ sum to zero. More problematic is an MHV
construction for the other gravity amplitudes. Already the first
not vanishing googly amplitude $\mathcal{M}(+,+,-,-,-)$ involves a
{\it non} holomorphic 4 vertex. The naive application of the MHV
prescription of \cite{MHV} to this amplitude seems to fail. In
particular the result is not covariant. It is not clear to us
whether this failure is due to the special features of gravity
(\textit{e.g.}, lack of conformal invariance) which may lead to
the non equivalence of connected and disconnected prescriptions.
If this were the case one should sum over both connected and
disconnected configurations in the corresponding string theory.
Another possibility would be that our off-shell extension needs to
be modified.

\section{A googly graviton amplitude}
\setcounter{equation}{0}

Starting from the observation that a closed string vertex operator
factorizes into the product of two open string vertices, Kawai,
Lewellen and Tye \cite{klt} were able to derive a set of formulas
relating closed string amplitudes to open string ones. In the
low-energy limit these formulas imply a similar factorization of
gravity amplitudes as products of two gauge theory amplitudes.

By direct use of the KLT relations it has therefore been possible
\cite{bgk} to obtain compact expressions for several tree-level
gravity amplitudes, which would have been much more difficult to
compute  diagrammatically, considering the complexity of
perturbative gravity. A nice review of this topic is given in \cite{bern}.

Following \cite{bgk} we denote the amplitude for $n$ external
gravitons with momenta $p_1,\ldots, p_n $ and helicities $h_1,
\ldots, h_n$ by $\mathcal{M}(1h_1,\ldots, nh_n)$. Similarly to the
gluon case, the amplitude vanishes if more than $n-2$ gravitons
have the same helicity. The first non trivial amplitude describes
the scattering of 2 gravitons with one helicity and $n-2$
gravitons with the opposite one. The amplitude with $q=2$ negative
helicity gravitons is called maximally helicity violating (MHV),
whereas the amplitude with $q=n-2$ negative helicity gravitons is
called ``googly''.

In spinor helicity formalism the momentum of a massless particle
is expressed in terms of a $(\frac{1}{2},0)$ and a $(0,\frac{1}{2})$
commuting spinors ("twistors"), $\la_a$ and
$\tilde{\lambda}_{\dot a}$ ($a,\dot a=1,2$)
\bea
 p_{a\dot a}=\lambda_{a}\tilde{\lambda}_{\dot
 a}.
 \label{17}
 \eea
Following custom we will use the abbreviated notation for the
contraction of two spinors $\LA ij \RA=
\epsilon_{ab}\lambda_{i}^{a}\lambda_j^b$ and $[ij]=\epsilon_{\dot
a\dot b}\lla_i^{\dot a}\lla_j^{\dot b}$.

The explicit expression in the MHV case of $n=5$, $q=2$ gravitons
is \cite{bgk}
\bea
 \mathcal{M}(1-,2-,3+,4+,5+)=-4i\left(8\pi
 G_N\right)^{\frac{3}{2}}\frac{\LA
 12\RA^8}{\prod_{i=1}^4\prod_{j=i+1}^5 \LA
 ij\RA}\mathcal{E}(1,2,3,4)
 \label{184}
 \eea
where $\mathcal{E}(1,2,3,4)=\frac{1}{4i}\left([12]\LA 23\RA[34]\LA
41\RA- \LA 12\RA [23]\LA 34\RA [41]\right)$. This amplitude is of
the form
\bea
 \mathcal{M}(1-,2-,3+,4+,5+)=\sum_{\alpha=1,2}R_{\alpha}(\la_i)
 P_{\alpha}(\lla_i)
 \label{185}
 \eea
where the $R$'s are rational functions and the $P$'s are
polynomials. Even though (\ref{185}) is not holomorphic in $\la$
as (\ref{05}), it splits in two parts, each of them displaying a
simple polynomial dependence on $\lla$. This generalizes to all
MHV gravity amplitudes. As already shown in \cite{witten}, the
twistor transform of
\bea
 A(\la_i,
 \lla_i)=i(2\pi)^4\dd\bigg{(}\PP\bigg{)}\mathcal{M}(1-,2-,3+,4+,5+)
 \label{186}
 \eea
yields \footnote{The twistor transform coincides with a Fourier
transform in signature $++--$, where $\la$ and $\lla$ are
independent and real. As far as tree-level amplitudes are
concerned one can always switch signatures by Wick rotation.}
\bea
 \tilde{A}(\la_i, \mu_i)&=&i\int
 d^4x\,\int\frac{d^2\tilde{\lambda}_1}{(2\pi)^2}\ldots
\frac{d^2\tilde{\lambda}_5}
 {(2\pi)^2}\,
 e^{i\sum_{i=1}^5\Lla_i(\mu_{i\dot{a}}+x_{a\dot{a}}\La_i)}\mathcal{M}(\la_i,
 \lla_i)
 \nonumber\\
 &=&i\sum_{\alpha=1,2}R_{\alpha}(\la_i)
 P_{\alpha}\left(i\frac{\partial}{\partial \mu_{i\dot
 a}}\right)\int d^4 x\,\prod_{i=1}^5\delta^2(\mu_{i\dot a}+x_{a
 \dot a }\La_i).
 \label{19}
 \eea
The twistor transformed amplitude is thus supported on a curve of
degree $d=1$ and genus $g=0$, via a polynomial in derivatives of
the delta function.

Now we move to the googly amplitude, which is obtained by
switching the $\la$'s and the $\lla$'s in (\ref{184}) \footnote{In
Lorentz signature this amounts to a parity transformation since
$\lla=\pm \bar{\la}$.} \bea
 \lefteqn{\mathcal{M}(1+, 2+, 3-, 4-,
 5-)=[\mathcal{M}(1-, 2-, 3+,
 4+, 5+)]^*=\sum_{\alpha=1,2}P^*_{\alpha}(\la_i) R^*_{\alpha}(\lla_i)
\nonumber}\\
 & &=(8\pi G_N)^{\frac{3}{2}}\bigg{(}
 \frac{\LA 12\RA \LA 34\RA [12]^8}{[12][13][15][24][25][34][35][45]}
 +
 \frac{\LA 23\RA \LA 41\RA
 [12]^8}{[13][14][15][23][24][25][35][45]}\bigg{)}.
 \label{20}
 \eea
This amplitude obeys for each $i=1,\ldots, 5$ a homogeneity
condition
\bea
 \bigg{(}\La_i\frac{\partial}{\partial
 \La_i}-\Lla_i\frac{\partial}{\partial
 \Lla_i}\bigg{)}\mathcal{M}=-2h_i \mathcal{M}
 \label{205}
 \eea
where $h_i=\pm 2$ is the helicity of the $i$-th graviton.

The transform to twistor space of
\bea
 A(\la_i,
 \lla_i)=i(2\pi)^4\dd\bigg{(}\PP\bigg{)}\mathcal{M}(1+,2+,3-,4-,5-)
 \label{21}
 \eea
would be
\bea
 \tilde{A}(\la_i,\mu_i)=i\sum_{\alpha=1,2}P^*_{\alpha}(\lambda_i)\int
 d^4x\,\int\frac{d^2\tilde{\lambda}_1}{(2\pi)^2}\ldots
 \frac{d^2\tilde{\lambda}_5}{(2\pi)^2}\,
 e^{i\sum_{i=1}^5\Lla_i(\mu_{i\dot{a}}+x_{a\dot{a}}\La_i)}
 R^*_{\alpha}(\tilde{\lambda}_i).
 \label{22}
 \eea
The homogeneity condition in twistor space reads
\bea
 \Big{(}\La_i\frac{\partial}{\partial \La_i}+\mu_{i\dot a}
 \frac{\partial}{\partial \mu_{i\dot a
 }}\Big{)}\tilde{A}=(-2h_i-2)\tilde{A}.
 \label{206}
 \eea
This can be obtained from (\ref{205}) by  replacing $\Lla$ with
$i\frac{\partial}{\partial \mu_{\dot a}}$ and
$-i\frac{\partial}{\partial \Lla}$ with $\mu_{\dot a}$.

According to (\ref{1}), we expect $\tilde{A}$ to be supported on a
$d=2$, $g=0$ curve in twistor space. Since the $\lla$ dependence
of (\ref{20}) is through rational functions, it is not easy to
perform explicitly the twistor transform and check this
conjecture. Witten proposed an alternative way to avoid this
cumbersome computation \cite{witten}. This method is based on the
introduction of operators which control if a set of given points
lies on a common curve embedded in twistor space. These operators
are algebraic in the ($\la, \mu$) space, while they are
differential once transformed back to the ($\la,\lla$) space.

The relevant operator for the $n=5$, $q=3$ case is \bea
K_{ijkl}=\epsilon_{IJKL} Z^I_i Z^J_j Z^K_k Z^L_l \label{23} \eea
where $Z^I_i$ are homogeneous coordinates in $CP^3$, namely
$Z^I_i=(\lambda^1_i,\lambda^2_i,\mu_{i 1},\mu_{i 2})$, for the
$i$-th graviton ($i=1,\ldots, 5$). To go to the ($\la, \lla$)
space, one simply replaces $\mu_{i \dot a}$ with
$-i\frac{\partial}{\partial \tilde{\lambda}^{\dot a}_i}$. We
introduce the notation
\bea
 \{ ij \}=\epsilon^{\dot a\dot
 b}\frac{\partial^2}{\partial \tilde{\lambda}^{\dot a}_i\partial
 \tilde{\lambda}^{\dot b}_j}.
 \label{24}
 \eea
The differential operator in ($\la,\lla$) space is thus expressed
as
\bea
 K_{ijkl}=\LA ij\RA\{kl\}-\LA ik\RA\{jl\}-\LA jl\RA\{ik\}+\LA
 il\RA\{jk\}+\LA kl\RA\{ij\}-\LA jk\RA\{li\}.
 \label{25}
 \eea

If the amplitude is supported on a $d=2$, $g=0$ curve through a
delta function, then one expects that $K_{ijkl}A(\la,\lla)=0$.
This is indeed what happens for the $n=5$, $q=3$ tree-level gluon
amplitude, as verified in \cite{witten}. What we are actually
going to prove for the graviton amplitude is that
\bea
 K_{ijkl}
 K_{i' j' k' l'} A=0.
 \label{26}
 \eea
This means that we still have a localization on a $d=2$, $g=0$
curve but now via a derivative of the delta function. This is
somewhat similar to what happens in the 1-loop gluon amplitude.

A useful simplification in checking (\ref{26}) is achieved by
using the manifest Poincar\'{e} invariance of both $K$ and
$A(\la,\lla)$. The Lorentz transformations are given by
$SL(2,R)\times SL(2,R)$, with the first $SL(2,R)$ acting on the
$\la$'s and the second one on the $\lla$'s. Translations act on
the $\mu$'s as $\mu_{i\dot a}\rightarrow \mu_{i\dot a}+x_{a \dot
a}\La_i$. It is thus possible to fix two points in twistor space
$Z_i$, $Z_j$ to convenient values: $\la_i$ and $\la_j$ can be
fixed by use of $SL(2,R)$ plus a scaling allowed by (\ref{206}),
whereas $\mu_{i\dot a}$ and $\mu_{j\dot a}$ are fixed by the
translations.

We can choose for example to fix $Z_3=(1,0,0,0)$ and $Z_4=(0,1,0,0)$.
This means
$\la_3=(1,0)$, $\la_4=(0,1)$ and $\mu_3=\mu_4=(0,0)$.
The delta function of momentum conservation enforces
\bea
 \Lla_3=-\sum_{j=1,2,5}\la^1_j\Lla_j \nonumber \\
 \Lla_4=-\sum_{j=1,2,5}\la^2_j\Lla_j.
 \label{30}
 \eea
By substituting (\ref{30}) in (\ref{20}) we obtain a ``fixed''
amplitude $A^{fix}$, which is function only of $\la_i, \lla_i$
with $i=1,2,5$. We find that the dependence of $A^{fix}$ on the
$\lla$'s is only through the bilinears $a\equiv[12]$,
$b\equiv[15]$ and $c\equiv [25]$. The crucial property of
$A^{fix}$ is that
\bea
 \left(a\frac{\partial}{\partial
 a}+b\frac{\partial}{\partial b}+c\frac{\partial}{\partial
 c}\right)A^{fix}=0.
 \label{31}
 \eea
This follows directly from the observation that the original
amplitude (\ref{20}) is homogeneous of degree 0 in the
antiholomorphic bilinears. Since (\ref{30}) is linearly
homogeneous in the $\lla$'s, the fixed amplitude is still
homogeneous of degree 0 in $a, b, c$.

After fixing $Z_3$ and $Z_4$, (\ref{25}) can also be expressed in
terms of $a$, $b$ and $c$. Defining an operator $\hat{O} \equiv
(a\frac{\partial}{\partial a}+b\frac{\partial}{\partial
b}+c\frac{\partial}{\partial c}+1)$ we find
\bea
 K_{1234}&=&-\frac{\partial}{\partial a} \hat{O} \nonumber \\
 K_{1345} &=& -\frac{\partial}{\partial b} \hat{O} \nonumber \\
 K_{2345} &=& -\frac{\partial}{\partial c} \hat{O} \nonumber \\
 K_{1235} &=& -\bigg{(}\la_5^2 \frac{\partial}{\partial a}- \la_2^2
 \frac{\partial}{\partial b}
 +\la_1^2 \frac{\partial}{\partial c}\bigg{)} \hat{O} \nonumber \\
 K_{1245} &=& - \bigg{(}-\la_5^1 \frac{\partial}{\partial a} +\la_2^1
 \frac{\partial}{\partial b}
 -\la_1^1 \frac{\partial}{\partial c}\bigg{)} \hat{O}.
 \label{32}
 \eea
These are the only independent operators up to permutations. Since
$A^{fix}$ is homogeneous of degree zero, $\hat{O}
A^{fix}=A^{fix}$, and it follows that no component of $K$
annihilates the amplitude.

However from (\ref{32}) it can be seen that $K_{ijkl}A^{fix}$ is
homogeneous of degree -1 in $a$,$b$, and $c$ for every $i,j,k,l$,
and thus it will be annihilated by the operator $\hat{O}$. From
this observation we conclude \bea
 K_{ijkl} K_{i' j' k' l'} A^{fix}
 = 0
 \label{33}
 \eea
for any choice of $i,j,k,l$ and $i' j' k' l'$.

\section{Disconnected MHV decomposition}
\setcounter{equation}{0}

So far we have investigated the possibility for a twistor
transformed gravity amplitude to be localized on connected curves
whose degree and genus are given by (\ref{1}). In the gauge theory
context of \cite{witten}, a certain string interpretation suggests
that also disconnected curves may play a role in the computation
of amplitudes, and that a connected contribution might be
decomposed into disconnected pieces. An amplitude supported on a
degree 2 curve can, for example, receive contributions from
configurations with two disconnected degree 1 curves. Although one
expects a contribution from all the possible decompositions, in
\cite{MHV} it was shown that tree-level gauge theory amplitudes
can be obtained by taking the completely disconnected
configuration only. Inspired by what happens in the gauge theory,
we try to check if a similar decomposition holds for gravity as
well.

In this Section we present the 3 and 4 graviton vertices given by
the $(+,-,-)$ and $(+,+,-,-)$ MHV amplitudes and we try to apply
this procedure to some simple gravity amplitudes, including the
$n=5$ googly one studied in Section 2.

\subsection{The $(+,-,-,-)$ and $(+,-,-,-,-)$ amplitudes}

\begin{figure}
\begin{center}
\includegraphics[width=145mm]{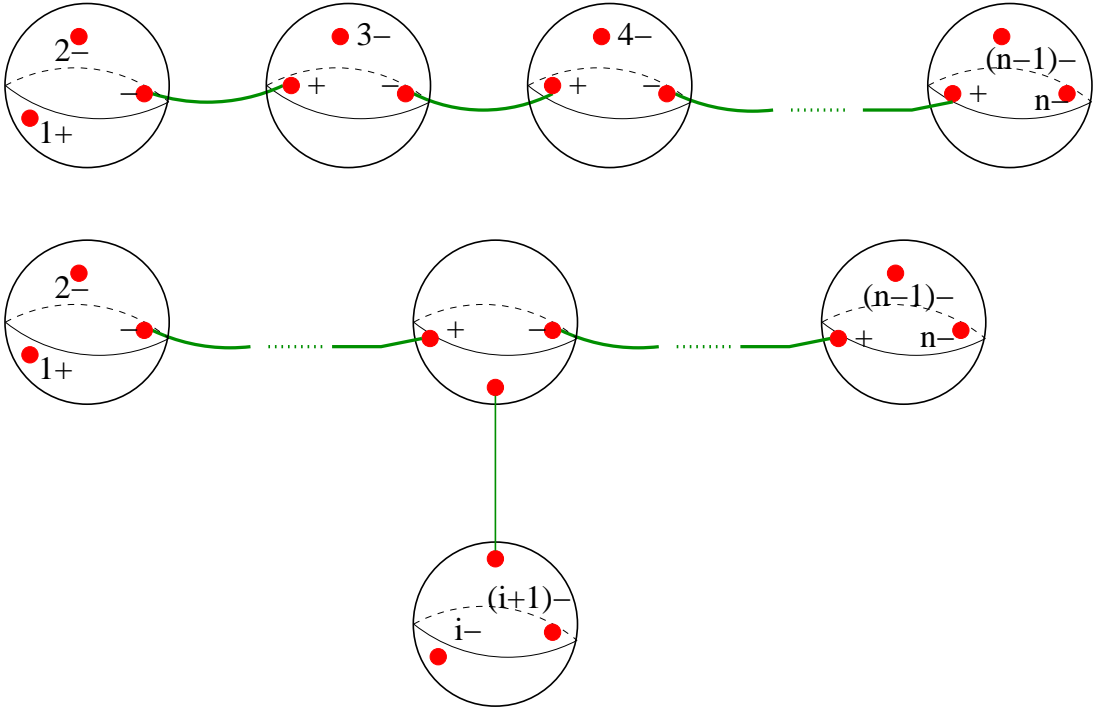}
\caption{Two disconnected configurations contributing to
$\mathcal{M}_n(1+,2-,\ldots,n-)$.}\label{fig1}
\end{center}
\end{figure}

\begin{figure}
\begin{center}
\includegraphics[width=85mm]{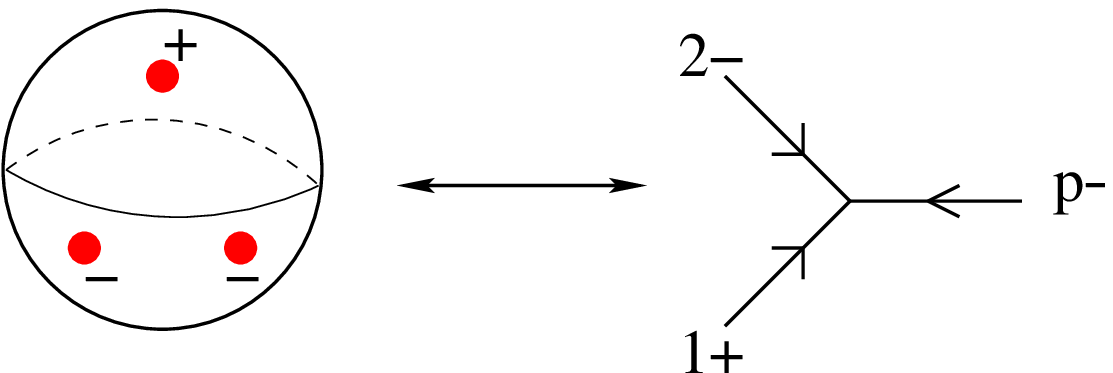}
\caption{The $(+,-,-)$ graviton vertex.}\label{fig2}
\end{center}
\end{figure}

\begin{figure}
\begin{center}
\includegraphics[width=145mm]{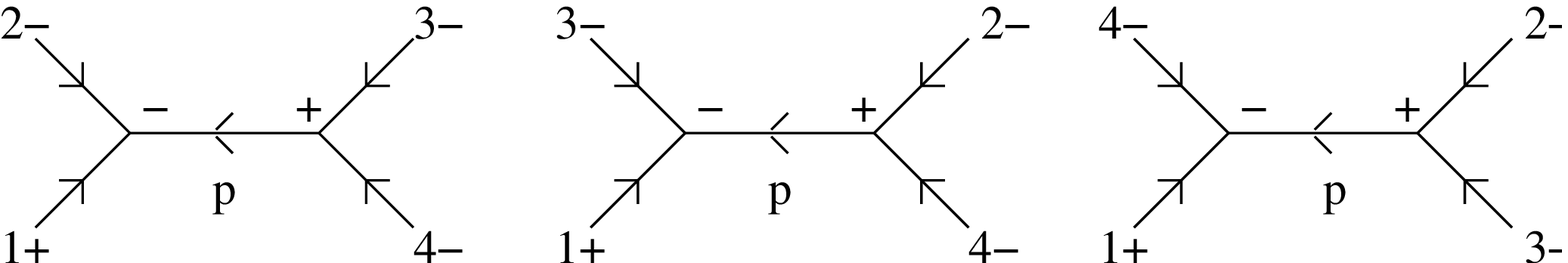}
\caption{The MHV diagrams contributing to the
 $\mathcal{M}(1+,2-,3-,4-)$ graviton amplitude.}\label{fig3}
\end{center}
\end{figure}

Amplitudes of the type $\mathcal{M}(1+,2-,\ldots,n-)$ should
correspond to the twistor space diagrams in Fig. \ref{fig1}. As
already stated, these are known to vanish.
Each $CP^{1}$ represents a $(+,-,-)$ vertex, Fig. \ref{fig2}.
This vertex is obtained by suitably extending the vanishing
$(+,-,-)$ graviton amplitude off-shell. This is formally given by
the square of the corresponding gluon amplitude \footnote{The
general KLT factorization formula relating closed and open string
amplitudes reads
$\mathcal{M}^{closed}_n\sim\sum_{p,p'}\mathcal{M}^{open}_n(p)
\tilde{\mathcal{M}}^{open}_n(p')e^{i\pi F(p,p')}$ where $p$ and
$p'$ are different orderings of the $n$ external legs. In the
$n=3$ case the phase factor $e^{i\pi F(p,p')}$ drops out yielding
$\mathcal{M}^{closed}_3\sim\mathcal{M}^{open}_3\tilde{\mathcal{M}}^{open}_3$.
In the $\alpha'\rightarrow 0$ limit this translates to a similar
relation between gravity and gauge theory amplitudes.} \cite{klt}.
The off-shell extension of the twistor $\la_p$ corresponding to an
off-shell momentum $p$ has been given in \cite{MHV} and amounts to
defining \bea
 \la_{pa}=\frac{p_{a\dot{a}}\eta^{\dot a}}{[\tilde{\la}_p,\eta]}
 \label{40}
 \eea
where $\eta^{\dot a}$ is an arbitrary spinor. The normalization factor
is needed
in order to have a consistent on-shell limit, and it can be dropped if
the amplitude is
homogeneous in the $\la_p$.  The off-shell
extension of the 3 graviton amplitude is therefore
 \bea
 \mathcal{M}_3=\left( \frac{\LA 2, p\RA^4}{\LA 1,2 \RA
 \LA 2,p \RA \LA p,1 \RA} \right)^2.
 \label{41}
  \eea
In this section we start focusing on  $\mathcal{M}(1+,2-,3-,4-)$.
This is computed using the MHV diagrams shown in Fig. \ref{fig3}.

\begin{figure}
\begin{center}
\includegraphics[width=145mm]{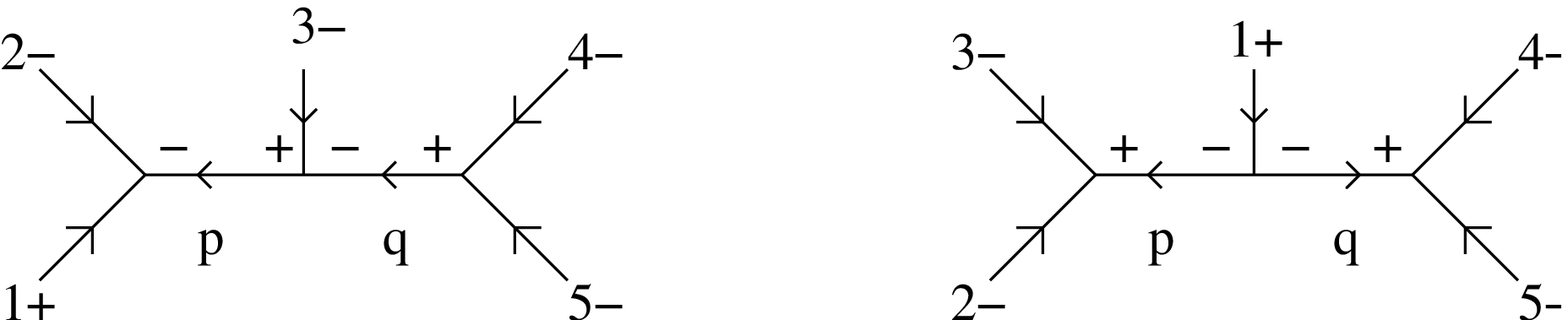}
\caption{Two of the fifteen MHV diagrams contributing to the
 $\mathcal{M}(1+,2-,3-,4-,5-)$ graviton amplitude.}\label{fig4}
\end{center}
\end{figure}
The contribution of the first graph is given by
\bea
 \frac{\LA 2p \RA ^8}{(\LA 12 \RA\LA 2p\RA \LA p1\RA)^2}
 \frac{1}{p^2} \frac{\LA 34\RA ^8}{(\LA p3\RA \LA 34\RA \LA
 4p\RA)^2}=\frac{\phi_1^6}{\phi_2^2 \phi_3^2 \phi_4^2}\frac{\LA
 12\RA\LA 34 \RA ^2}{[12]}
 \label{42}
 \eea
where we have used
$\la_{pa}=-\la_{1a}\phi_1-\la_{2a}\phi_2=\la_{3a}\phi_3+\la_{4a}\phi_4$,
with $\phi_i=\tilde{\la}_{i\dot a}\eta^{\dot a}$. The remaining
two diagrams are obtained by appropriately permuting the external
labels. Using momentum conservation in the form of $\sum_{i=1}^4
\LA yi \RA [iz]=0$ (where $\la_y$ and $\lla_z$ are arbitrary
spinors), the final result can be arranged as
\bea
 \mathcal{M}(1+,2-,3-,4-)=\frac{\phi_1^6}
 {\phi_2^2\phi_3^2\phi_4^2}\bigg(\LA 12 \RA
 \LA 34\RA + \LA 13 \RA
 \LA 42\RA + \LA 14 \RA
 \LA 23\RA\bigg{)} \frac{\LA 42 \RA}{[13]}.
 \label{43}
 \eea
This vanishes by virtue of the Schouten identity $\LA ij\RA\LA
kl\RA+\LA ik\RA\LA lj\RA+\LA il\RA\LA jk\RA=0$ which is valid for
any four spinors.

Moving now to $\mathcal{M}(1+,2-,3-,4-,5-)$ we need to consider
graphs of the type given in Fig. \ref{fig4}. The first diagram
gives
\bea
 \frac{\phi_1^6}{\phi_2^2\phi_3^2\phi_4^2\phi_5^2}
 \frac{\LA 12\RA\LA 45\RA(\LA 34\RA \phi_4 +\LA 35\RA
 \phi_5)^6}{[12][45](\LA 13\RA \phi_1 +\LA 23\RA \phi_2)^4}
 \label{44}
 \eea
where we have extended both $\la_{pa}$ and $\la_{qa}$ off-shell
using the same spinor $\eta^{\dot a}$. This diagram yields 12
contributions once one takes into account all inequivalent
exchanges of the negative helicity external gravitons. The second
graph gives
\bea
 \frac{\phi_1^6}{\phi_2^2\phi_3^2\phi_4^2\phi_5^2}
 \frac{\LA 23\RA \LA
 45\RA (\LA 12 \RA \phi_2 +\LA 13\RA \phi_3)^4}{[23][45](\LA 14\RA
 \phi_4+\LA 15\RA \phi_5)^2}
 \label{45}
 \eea
and 2 other terms obtained by permutations. Imposing momentum
conservation, with some computer assistance
 one can verify that the sum of the 12 contributions
coming from (\ref{44}) and the 3 contributions coming from
(\ref{45}) vanishes as expected.

We stress here the holomorphicity of (\ref{41}), which is the only
vertex appearing in this kind of graphs.

\subsection{The googly amplitude}

\begin{figure}
\begin{center}
\includegraphics[width=35mm]{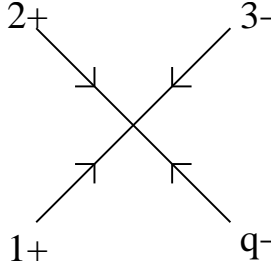}
\caption{The $(+,+,-,-)$ graviton vertex.}\label{fig5}
\end{center}
\end{figure}

\begin{figure}
\begin{center}
\includegraphics[width=145mm]{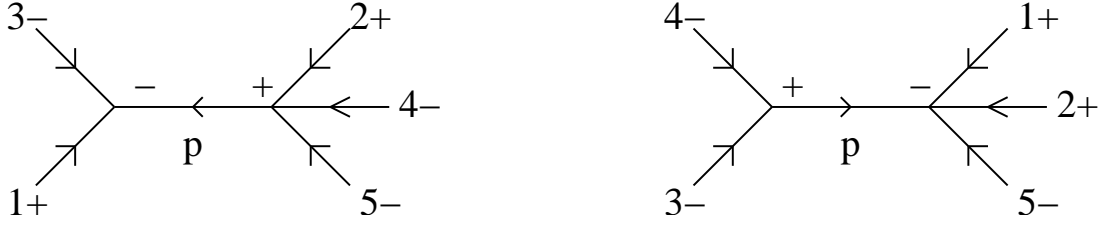}
\caption{Two of the nine MHV diagrams contributing to the
 $\mathcal{M}(1+,2+,3-,4-,5-)$ graviton amplitude.}\label{fig6}
\end{center}
\end{figure}

We now come to the investigation of disconnected contribution to
$\mathcal{M}(1+,2+,3-,4-,5-)$. In the construction of the MHV
graphs one also needs here the 4 graviton vertex depicted in Fig.
\ref{fig5}.
The expression for the 4 graviton amplitude was first obtained in
\cite{bgk} and is given by
\bea
 \mathcal{M}(1+,2+,3-,q-)=\frac{\LA
 3q\RA^8}{\LA 12\RA\LA 13\RA\LA 1q\RA\LA 23\RA\LA 2q\RA\LA 3q\RA}
 \frac{[3q]}{\LA 12\RA}.
 \label{46}
 \eea
One immediately notices that this expression is not holomorphic
and this is in strong contrast with the 3 graviton vertex
(\ref{41}) and all the gluon MHV vertices. Naively, an off-shell
extension of (\ref{46}) would require a redefinition of $\Lla$
whenever it appears in an internal line. Hermiticity suggests to
take the complex conjugate of (\ref{40}) so to have
\bea
 \tilde{\la}_{p\dot a}=\frac{p_{a\dot{a}}\xi^{a}}{\LA\la_p,\xi\RA}
 \label{47}
 \eea
where $\xi=\eta^*$. Using this prescription one gets for the first
graph in Fig. \ref{fig6}
\bea
 \frac{\phi_1^6}{\phi_3^2}
 \frac{\LA 13\RA \LA 45\RA ^7 [45]}{\LA 25\RA \LA 24\RA [13]
 (\LA 25 \RA \phi_2 -\LA 54\RA \phi_4)(\LA 24 \RA \phi_2+ \LA 54\RA \phi_4)
 (\LA 25\RA \phi_5+\LA 24\RA \phi_4)^2}
 \label{48}
 \eea
and for the second graph
\bea
 \frac{1}{\phi_3^2\phi_4^2 (\phi_3 \tilde{\phi_3}+\phi_4 \tilde{\phi_4})}
 \frac{\LA 34\RA (\LA 15\RA \phi_1 +\LA 25 \RA\phi_2)^7([15]\tilde{\phi}_1
 +[25]\tilde{\phi}_2)}{\LA 15\RA \LA 25\RA \LA 12\RA^2[34](\LA 12\RA \phi_2
 +\LA 15\RA \phi_5)(\LA 25\RA \phi_5-\LA 12\RA \phi_1)}
 \label{49}
 \eea
where $\tilde{\phi}_i=\lambda_{i a}\xi^a$. The factor
$\phi_3\tilde{\phi_3}+\phi_4\tilde{\phi_4}=[\lla_p, \eta] \LA \la_p,
\xi\RA$ comes from the
normalization of (\ref{40}) and (\ref{47}) which does not cancel in this case.
One can get all the
other seven graphs by permutation of the external labels as usual.
The expected result for this amplitude is given in (\ref{20}),
which some computer algebra showed not to match with the one
following from (\ref{48}) and (\ref{49}). Moreover, the result depends
on $\eta$.
Therefore the prescription seems to
fail in this case.
We are aware of the fact that the heuristic proof of covariance
given in \cite{MHV}
might not be generalizable in the presence of non holomorphic vertices.

\section{Conclusion}
\setcounter{equation}{0}

In this note we have explored the possibility of extrapolating the
twistor construction of \cite{witten} to ordinary gravity. We have
checked that the simplest non-trivial gravity quantity, namely the
5 graviton googly amplitude, confirms the expectations of \cite{witten}
and is indeed supported on a connected degree 2 curve in twistor
space, just as the corresponding amplitude in the gauge theory
\footnote{The computation does not exclude additional contributions
coming from disconnected, lower degree curves.}. There are however
important differences between the two. In the simplest, MHV case,
these stem from the fact that gravity amplitudes contain extra
delta-function derivatives in twistor space variables, or equivalently
they are not holomorphic in Minkowski space variables. It is clearly
desirable to confirm that such behavior persists for further,
non MHV graviton amplitudes.

In a complementary approach to the computation presented in
Section 2, we have further tried calculating tree-level graviton
amplitudes by using MHV subamplitudes as vertices (computed from
the gauge theory quantities by using the KLT relations, and
suitably continuing them off-shell), in the spirit of the
prescription given in \cite{MHV} for gauge theories. Although it
is possible that such a generalization might be feasible in
principle, it is clear from our results that novel ingredients are
necessary to correctly reproduce non-trivial gravity amplitudes.

We nevertheless consider it encouraging that the $(+,-,-,-)$ and
$(+,-,-,-,-)$ graviton amplitudes vanish when computed from MHV
vertices. We are aware that these are very special cases. Indeed,
$(+,-,\ldots,-)$ amplitudes involve only trivalent MHV vertices,
which are holomorphic even in the graviton case. Unfortunately, the
four-valent graviton MHV vertex is not holomorphic. We believe that
this non-holomorphicity is an important reason for the failure of
the MHV prescription to correctly reproduce the 5 graviton googly
amplitude discussed in this note.

We must emphasize that the twistor string theory underlying an
eventually successful version of such a construction might have
nothing to do with the one of \cite{witten}, or even there might be no
such theory at all. Indeed, the closed string sector of the model
of \cite{witten} is expected to be a kind of instanton expansion
around ${\cal N}=4$ self-dual superconformal gravity. General
Relativity is most definitely not conformally invariant, and therefore
 it should be related to a different model. The
computation in Section 2 seems to suggest that there could be some localization
 in twistor space, and the disconnected
prescription could provide an explicit and computable
``instanton'' expansion around some ``self-dual'' theory. In this
respect, we think that the non-holomorphicity of higher MHV
vertices could provide a hint about which could be the right
theory to expand around.


\vskip 15mm
\begin{center}\subsubsection*{Acknowledgments}\end{center}

We would like to thank Zvi Bern, Freddy Cachazo, Martin Ro\v{c}ek,
and Warren Siegel for helpful discussions. The research of R.R. and D.R.-L. is partially supported by NSF Grant No. PHY0098527.



\end{document}